# Photometric observations of ecliptic comet 47P/Ashbrook-Jackson and selected quasi-Hilda and main-belt comets at Kyiv Comet Station (MPC code – 585) in 2017

Serhii Borysenko[1], Alexander Baransky[2], Elena Musiichuk[3]

[1]Main Astronomical Observatory of NAS of Ukraine, Akademika Zabolotnoho 27, Kyiv 03143, Ukraine

e-mail: borisenk@mao.kiev.ua

[2]Astronomical Observatory of Taras Shevchenko National University of Kyiv Observatorna str. 3, Kyiv, 04053 Ukraine

[3]Taras Shevchenko Kyiv University, Glushkova Ave. 2, Kyiv 03022, Ukraine

Pages: 14

Tables: 2

Figures: 1

**Proposed Running Head**: Photometric observations of ecliptic comet 47P/Ashbrook-Jackson and selected quasi-Hilda and main-belt comets at Kyiv Comet Station (MPC code – 585) in 2017

**Editorial correspondence to:**

Dr. Serhii Borysenko, Ph.D.

Main Astronomical Observatory of NAS of Ukraine

 Akademika Zabolotnoho str. 27

Kyiv, 03143

Ukraine

Phone: +380 44 526-47-69

E-Mail address: borisenk@mao.kiev.ua

**ABSTRACT**

We present an analysis of the photometric data of comets *47P/Ashbrook – Jackson*, *65P/Gunn, 362P/2008 GO98,* and *P/2017 S5 (ATLAS)* observed at Kyiv Comet Station (MPC code – 585) in 2017. The upper limits of radii for cometary nuclei, *Afρ* parameters, and color-indexes are measured.

**Key words:** Comet; Photometry; Dust productivity; Radius of the nucleus; Color-index

1. Introduction

There are two big groups of comets with orbits between 2 – 5 AU. Cometary objects with orbits in the outer part of the main asteroid belt are known as main-belt comets (MBCs). Dynamical studies suggest that they should be formed in-situ beyond the primordial snow line. Comets formed at such distances from the Sun represent a class of icy bodies which has not yet been studied in detail. It could potentially hold the key to understanding the origin of water on terrestrial habitable worlds (Hsieh and Jewitt, 2006; Hsieh et al., 2011; Bertini, 2011; Jewitt, 2012). Comets in this group can have only a small coma, but they have the rich morphology of activity (short or long direct tails, jets and so on) (Jewitt et al., 2013; Jewitt et al. 2015a).

Another group of cometary objects is associated with the Hilda asteroid zone inside the 3:2 mean-motion resonance with Jupiter, so-called quasi-Hilda comets (QHCs). Usually comets in this zone have semimajor axes between 3.7 and 4.2 AU, eccentricities up to about 0.3, and inclinations of no more than 20° (Toth, 2006; Ohtsuka, 2008).

Small telescopes are still useful for broadband photometry of comets without emission lines in their spectra. The *R* pass-band covers the spectral range where usually comets display weak emissions, and the continuum dominates. For this reason, filters of this spectral range are suggested for comets' imaging and photometry related to the Afρ quantity. As a rule, MBCs and QHCs do not show relevant emission in the *R* band and other bands of the spectrum, in particular, in the *V* band (Jewitt et al., 2015b, Jewitt et al., 2009; Licandro et al., 2011).

## 2. Observations and analysis

In 2017, observations of selected comets were performed at the Kyiv Comet Station (MPC code – 585) of the Astronomical Observatory of Taras Shevchenko National University of Kyiv at Lisnyky village, Ukraine. Broadband Johnson filters *R* and *V* were used with FLI PL 47-10 CCD camera at the prime focus (F = 2800 mm) of the 0.7–m reflector. The detector consists of a 1024x1024 pixel matrix with a pitch of 13 $\mu m$ which corresponds to a scale of 0.947″ per pixel. The typical readout noise is about 10 $e^-$ and the conversion gain 1.22 $e^-$/ADU.

A log of observations is listed in Table 1.

[Table 1]

Preprocessing of the images was made with the IDL 8.4 software, while stacking of frames was made with *Astrometrica 4.0* (http://www.astrometrica.at/).

We used ATV IDL (http://www.exelisvis.com/ProductsServices/IDL.aspx, (Landsman, 1993)) routine (ver. 3.0b7) for aperture photometry of comets and reference stars. Most of the cometary images were compact with a tiny coma and a short tail. To obtain nuclear magnitudes of comets we used a method proposed by Hicks et al. (2007). APASS-9 star catalog was used as a reference. This catalog includes magnitudes of stars from about 7th magnitude to about 17th magnitude in five filters: Johnson *B* and *V*, plus Sloan *g′*, *r′*, *i′*. The mean uncertainties of catalog data are of about 0.07 mag for *B*, about 0.05 mag for *V* and less than 0.03 mag for *r′* (Henden et al., 2016). We used transformation formula from *r′* to *R* magnitudes suggested by (Munari et al., 2014). About 6 – 8 reference stars of 14th – 17th magnitude were used for each night. The main source of photometric errors is usually the signal-to-noise ratio (*SNR*) (Mikuz and Dintinjana, 1994). For most of images *SNR* was more than 20. The *Afρ* parameter (A'Hearn et al., 1984) was calculated as:

$$Af\rho = \frac{4r^2 \Delta^2 \cdot 10^{0.4(m_\odot - m_c)}}{\rho} \quad (1)$$

Here, r [AU] is the heliocentric distance; Δ [cm] is the geocentric distance; $m_\odot$ and $m_c$ are the apparent *R* magnitudes of the Sun and a comet, respectively; ρ [cm] is the radius of the photometric aperture projected onto the sky. A rough estimation of upper limits of radius for cometary nuclei was calculated as:

$$R_n^2 = \frac{2.238 \cdot 10^{22} r^2 \Delta^2 10^{0.4(m_\odot - m_{nucl})}}{p_R \cdot 10^{-0.4\alpha\beta}} \quad (2)$$

Here, r [AU] is the heliocentric distance; Δ [AU] is the geocentric distance; $m_\odot$, $m_{nucl}$ are the apparent *R* magnitude of the Sun and a nuclear magnitude of a comet respectively; α [deg] is the phase angle and β = 0.04 [mag/deg] - phase coefficient (Russel, 1916; Jewitt, 2009). For the geometric albedo in the *R* filter we decided to take the classical value of $p_R$ = 0.04 for the quasi-Hilda comets (as for Jupiter-family comets) and $p_R$ = 0.05 for the main-belt comet *P/2017 S5 (ATLAS)* (as for C-type asteroids) (Hsieh et al., 2009). The solar *R* magnitude used is $m_\odot$ = –27.29 mag (Cox, 2000). Samples of images are presented in Fig.1.

[Fig 1]

A log of observations and results of measurements are listed in Table 1.

[Table 1]

The aperture radii of apparent total red magnitude estimations were 8″ – 16″ (15 243 – 23 201 km) for 47P comet, 20″ – 12″ (31 064 – 19 710 km) for 65P comet, 16″ – 9″ (28 023 – 18 142 km) for 362P comet and 10″ (9 164 – 10 507 km) for P/2017 S5 comet during of the observational periods.

## 3. Objects of interest

*47P/Ashbrook-Jackson* was independently discovered by astronomers in the United States and South Africa. The first discovery was made by Yale astronomer Joseph Ashbrook. The comet's orbital period is 8.38 years (*a* = 4.13, *e* = 0.32). It has been observed at every return following the 1948 discovery. Comet 47P/Ashbrook-Jackson has Tisserand parameter with respect to Jupiter of $T_J$=2.90, that is significantly less than 3.00, i.e., the upper limit for Jupiter-family comets (A'Hearn et al., 1995).

*65P/Gunn* is a periodic comet in the Solar System which has a current orbital period of 6.79 years (*a* = 3.88 AU, *e* = 0.25). It was discovered on October 11, 1970, by Professor James E. Gunn of Princeton University using the 1.22-m Schmidt telescope at the Palomar Observatory. The comet is a short-period comet, orbiting the Sun every 6.79 years between the orbits of Mars and Jupiter. Wide-field Infrared Survey Explorer (WISE) observed the comet on April 24, 2010, just one month after the comet's closest approach to the Sun. It looked something like a swordfish or narwhal. This "sword", or dust trail, is made of dust particles that have previously been shed by *65P/Gunn* as it orbits the Sun. The dust is warmed by sunlight and glows in infrared light. Trails appear both ahead and behind the comet's nucleus and have a narrow, contrail-like appearance (http://www.jpl.nasa.gov/spaceimages/index.php, http://www.ast.cam.ac.uk/jds/per01.htm).

*362P/2008 GO98* is located in an unstable orbit (*a* = 3.96, *e* = 0.28) that takes it near Jupiter every few decades. It is currently ranked among the asteroids of the outer main-belt, specifically among the asteroids of the Hilda family. However, the discovery of a coma and a tail raises the question of its true nature. The cause of this activity is not known: it could be an asteroid become active following the collision with another small body that is not known for its part, or a sleeping comet that would have woken up. It could, therefore, be a quasi-Hilda comet. R. Gil-Hutton and E. García-Migani include this object among the quasi-Hilda objects and list it as a potential comet (Gil-Hutton and Garcia-Migani, 2016).

*P/2017 S5 (ATLAS)* was discovered on CCD exposures taken on 2017 Sept. 27.5 UT with the ATLAS 0.5-m f/2.0 Schmidt telescope at Haleakala in the course of the "Asteroid Terrestrial-impact Last Alert System" (ATLAS) search program. The orbit of the comet is located in the outer main asteroid belt ($a = 3.17$, $e = 0.31$). Reporting on the discovery, Aren Heinze noted that the object showed clear cometary features with a tail about 20" long toward position angle approximately 235 degrees, adding that the head of the comet also seemed elongated in the corresponding northeast-southwest direction. Pre-discovery ATLAS observations were subsequently identified from four additional nights between Sept. 11 and 23 (Heinze, 2017).

## 4. Discussion and conclusions

We present photometric observations of one ecliptic comet, of two QHCs and of a new main-belt comet *P/2017 S5 (ATLAS)* obtained during 2017. The *R* band images for all comets showed small comas elongated opposite the direction of the object motion (Figure 1). Most of observed comets show a color-index *V - R* about 0.5 - 0.6 (Table 2). Only the main-belt comet *P/2017 S5 (ATLAS)* show notable instability of color, which cannot be explained by measurement errors (Jewitt, 2015). Usually, short-term changes in V-R are consistent with injecting of fresh nuclear material into the coma. Some reddening in November, 15 can be explained by presence of additional dust in the cometary atmosphere, as a result of previous outburst (so as *R* magnitude and *Afρ* didn't increase).

**[Table 2]**

*47P/Ashbrook-Jackson* has increased its brightness few months after perihelion passage in June 2017. Maximum of *Afρ* index agrees with the maximum of brightness in the autumn. The increase in comet brightness is a result of its approach to the Earth (Table 1).

*65P/Gunn* still has been active during the summer. Its maximum of *Afρ* value in late May could be caused by an outburst with increasing of *R* magnitude. The coma profile was some diffuse that time. After the high activity in early 2016, the comet *362P/2008 GO98* slowly reduced its brightness during 2017. Typical values of *Afρ* for both QHCs varied from several tens to several hundred.

The new comet *P/2017 S5 (ATLAS)* showed *Afρ* value about 10 or less, which is typical for all MBCs (Hsieh, 2014).

Averaged profile brightness varied proportionally to distance from the photometric nucleus of comet as $\rho^{-0.75 \div -0.95}$ for 47P comet; as $\rho^{-0.60 \div -0.78}$ for 65P comet; as $\rho^{-0.63 \div -0.74}$ for 362P comet; as $\rho^{-0.93 \div -1.10}$ for P/2017 S5. The *1/ρ* brightness profile in the inner coma is consistent with a steady state dust outflow model, where dust leaves the nucleus isotropically at a constant speed, suggesting that there is active sublimation of volatiles driving the dust activity. More compact profiles for the comet 47P and P/2017 S5 can be as result of more solar radiation pressure on the smaller heliocentric distances (Table 1) (Jewitt and Meech, 1987).

For all observed comets $Af\rho$ values increased with the distance from nucleus $\rho$ for inner coma (up to 4″ – 6″) and then slowly decreased.

Estimations of the radii for cometary nuclei were made using a coma compensation method (Hicks et al., 2007). A stellar PSF obtained from background stars was used to fit the innermost coma. For cometary observations with dominated coma, we emphasize that the nuclear size estimates are upper limits only. First of all, it concerns the 65P comet with a conspicuous coma. For the analysis, we used only dates with the minimal activity of each comet for size estimations (Table 1) (Lamy et al., 2004; Lamy et al., 2011; Scotti, 1994).

Using the nuclear albedo of A=0.04, Lamy et al., 2004 and Lowry & Fitzsimmons, 2001; 2003 reported around 1 km smaller radius for 47P and 65P comets, they were 3.1 km and 8.8 km, respectively. Such differences can be explained by somewhat higher activity of the comets in 2017. Also we need take into account the fact that the nucleus of 47P is elongated. From other side, resolution about 1″/pix can be insufficient to obtain more realistic results.

**Table 1**

Log of *R* band observations and results of measurements

| Comet | Date | $N_i \times Exp$[a]. | $r$[b], AU | $\Delta$[c], AU | $\alpha$[d], deg | $R$[e] | $Af\rho$[f], cm | $m_{nucl}$[g] | $R_n$[h], km |
|---|---|---|---|---|---|---|---|---|---|
| 47P | 2017-07-18 | 11x60s | 2.827 | 2.624 | 21.1 | 17.59±0.04 | 50.9±1.9 | | |
| | 2017-08-10 | 16x60s | 2.840 | 2.347 | 19.8 | 17.35±0.07 | 43.0±2.8 | | |
| | 2017-08-11 | 18x60s | 2.841 | 2.335 | 19.6 | 17.22±0.10 | 48.2±4.4 | | |
| | 2017-09-12 | 15x60s | 2.869 | 2.030 | 13.2 | 16.58±0.05 | 41.1±1.9 | 18.21±0.02 | 4.4±0.1 |
| | 2017-09-14 | 15x60s | 2.872 | 2.010 | 12.4 | 16.40±0.03 | 45.1±1.2 | | |
| | 2017-09-16 | 19x60s | 2.875 | 1.997 | 11.8 | 16.21±0.05 | 53.5±2.5 | | |
| | 2017-09-27 | 20x60s | 2.888 | 1.944 | 8.2 | 16.27±0.02 | 53.1±1.0 | | |
| 65P | 2017-04-26 | 4x60s | 3.035 | 2.139 | 10.3 | 14.11±0.03 | 354±10 | | |
| | 2017-05-04 | 8x60s | 3.025 | 2.076 | 7.8 | 14.81±0.03 | 179±5 | | |
| | 2017-05-15 | 10x60s | 3.009 | 2.010 | 3.6 | 14.74±0.02 | 183±3 | | |
| | 2017-05-29 | 7x60s | 2.993 | 1.982 | 1.8 | 13.93±0.06 | 251±14 | | |
| | 2017-06-01 | 12x60s | 2.989 | 1.983 | 2.9 | 14.84±0.02 | 216±4 | | |
| | 2017-07-17 | 12x60s | 2.946 | 2.262 | 16.7 | 15.35±0.03 | 141±4 | 16.93±0.02 | 9.7±0.1 |
| 362P | 2017-07-18 | 18x60s | 3.312 | 2.412 | 9.6 | 16.30±0.03 | 63.1±1.7 | | |
| | 2017-07-19 | 20x30s | 3.314 | 2.408 | 9.3 | 16.83±0.05 | 55.4±2.6 | | |
| | 2017-08-11 | 15x60s | 3.369 | 2.390 | 5.3 | 16.56±0.03 | 85.0±2.3 | | |
| | 2017-09-14 | 13x60s | 3.452 | 2.617 | 10.7 | 17.46±0.06 | 51.2±2.8 | 19.74±0.04 | 3.2±0.1 |
| | 2017-09-27 | 18x60s | 3.485 | 2.776 | 13.1 | 17.53±0.05 | 57.6±2.7 | | |
| P/2017 S5 | 2017-09-29 | 26x60s | 2.217 | 1.262 | 10.3 | 17.39±0.03 | 10.9±0.3 | | |
| | 2017-11-15 | 23x60s | 2.292 | 1.447 | 16.2 | 17.80±0.05 | 9.1±0.3 | 18.50±0.02 | 2.1±0.1 |

[a] Number of stacking images and the exposure of the each one;
[b] Heliocentric distance;
[c] Geocentric distance;
[d] Phase angle;
[e] Apparent total red magnitude of the comet;
[f] $Af\rho$ parameter of the comet;
[g] Apparent nuclear red magnitude of the comet;
[h] Upper limit of radius for the cometary nucleus;

**Table 2**

Log of *V* band observations and results of measurements

| Comet | Date | $N_i \times Exp^a$. | $V^b$ | V - R |
|---|---|---|---|---|
| 47P | 2017-09-14 | 20x60s | 16.95±0.03 | 0.55±0.04 |
|  | 2017-09-27 | 27x60s | 16.86±0.03 | 0.59±0.04 |
| 65P | 2017-07-17 | 20x60s | 15.89±0.05 | 0.54±0.06 |
| 362P | 2017-07-18 | 20x60s | 16.86±0.03 | 0.56±0.04 |
|  | 2017-07-19 | 20x30s | 17.40±0.09 | 0.57±0.10 |
|  | 2017-08-11 | 9x60s | 17.17±0.08 | 0.55±0.06 |
|  | 2017-09-14 | 9x60s | 18.22±0.11 | 0.76±0.13 |
| P/2017 S5 | 2017-09-29 | 31x60s | 17.82±0.04 | 0.43±0.05 |
|  | 2017-11-15 | 10x60s | 18.62±0.07 | 0.82±0.09 |

[a] Number of stacking images and exposure of the each one;
[b] Apparent total visual magnitude of the comet;

**Fig 1**

Samples of the obtained images: a) 47P/Ashbrook-Jackson (2017-07-18), b) 65P/Gunn (2017-04-26), c) 362P/2008 GO98 (2017-07-18), d) P/2017 S5 (2017-09-29).

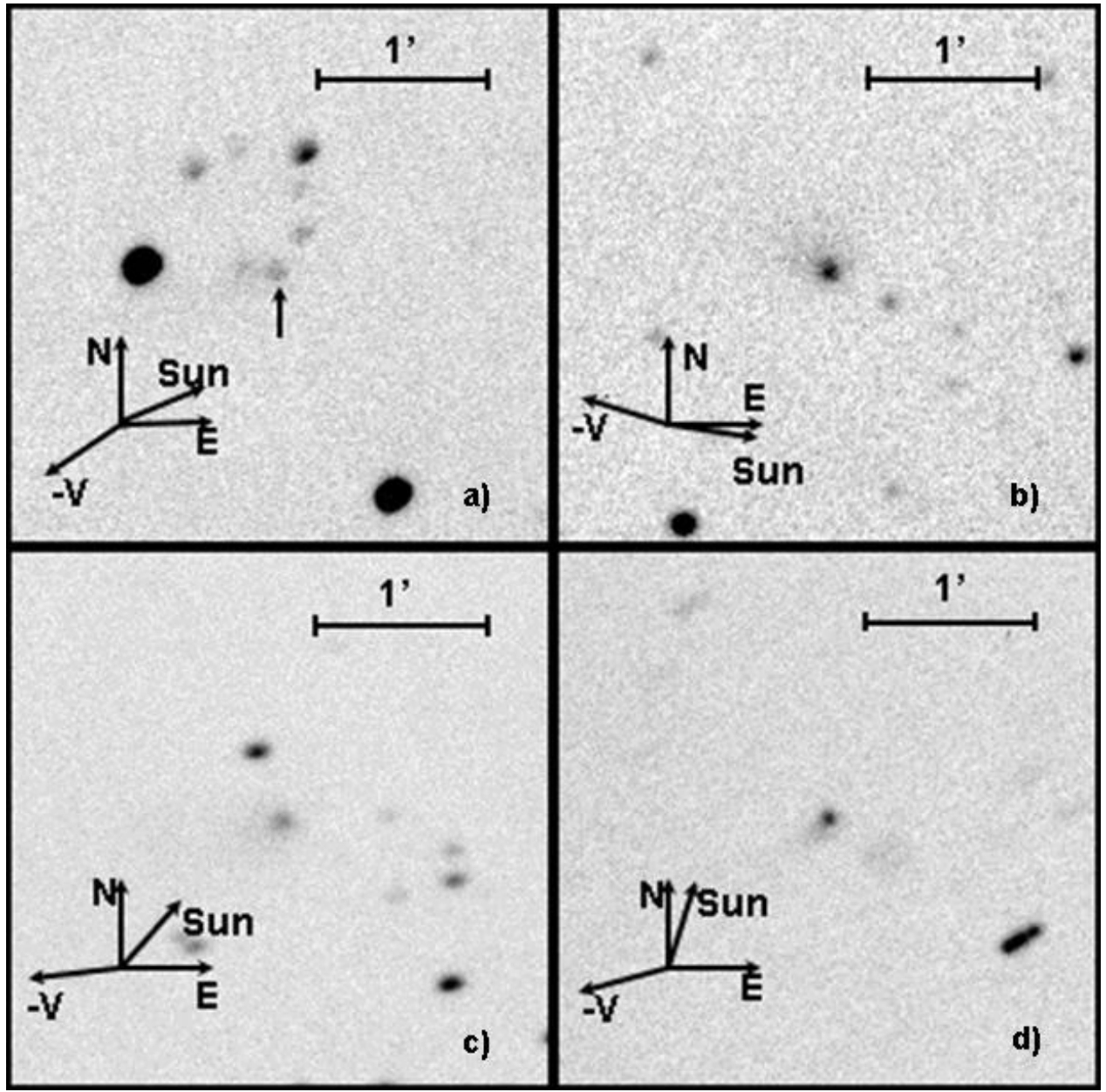